%% file: ms.tex
\documentclass[letterpaper, 10 pt, conference]{ieeeconf}  

\usepackage[style=ieee, backend=bibtex]{biblatex}
\input{packages.tex}

\input{colordef.tex}

\input{commands.tex}

\input{theorems.tex}

\input{tikzdef.tex}   

\makeatletter
\newcommand\notsotiny{\@setfontsize\notsotiny\@vipt\@viipt}
\makeatother

\IEEEoverridecommandlockouts                              

\title{\LARGE \bf
Time-optimal control with direct collocation and variable discretization
}

\author{Christoph R\"osmann, Artemi Makarow and Torsten Bertram
\thanks{This work was supported by the German Research Foundation (DFG, \mbox{BE 1569/13-1}). %
\textit{(Corresponding author: Christoph R\"osmann.)}}
\thanks{The authors are with the Institute of Control Theory and Systems Engineering, TU Dortmund University, 44227 Dortmund, Germany
        {\tt\small \{forename.surname\}@tu-dortmund.de}}%
}

\begin{document}

\maketitle
\thispagestyle{empty}
\pagestyle{empty}

\begin{abstract}

This paper deals with time-optimal control of nonlinear continuous-time systems based on direct collocation. 
The underlying discretization grid is variable in time, as the time intervals are subject to optimization.
This technique differs from approaches that are usually based on a time transformation.
Hermite-Simpson collocation is selected as common representative in the field of optimal control and trajectory optimization.
Hereby, quadratic splines approximate the system dynamics. Several splines of different order are suitable for the control parameterization.
A comparative analysis reveals that increasing the degrees of freedom in control,  e.g. quadratic splines, is not suitable for time-optimal control problems due to constraint violation and inherent oscillations.
However, choosing constant or linear control splines points out to be very effective. A major advantage is that the implicit solution of the system dynamics is suited for stiff systems and often requires smaller grid sizes in practice.
\end{abstract}

\input{introduction.tex}

\input{theory.tex}
\input{example.tex}
\input{summary.tex}

\addtolength{\textheight}{0cm}   


\printbibliography

\end{document}

%% file: packages.tex
\usepackage[super]{nth} 

\usepackage{amsmath}%
\usepackage{amsfonts}%
\usepackage{amssymb}%
\usepackage{mathtools}%

\usepackage{mathrsfs} 
\usepackage{nccmath}

\usepackage[style=american]{csquotes}

\usepackage[exponent-product = \cdot,
			output-complex-root = j, 
			separate-uncertainty = true,
			output-product = \cdot,
			arc-separator = \,,
			product-units = brackets-power]{siunitx}
\usepackage{subcaption}

\usepackage{tikz,pgfplots}

\usepackage{varwidth} 

%% file: colordef.tex
\definecolor{black}{rgb}{0.0,0.0,0.0}
\definecolor{darkgray}{rgb}{0.48,0.48,0.48}
\definecolor{darkgraytext}{rgb}{0.32,0.32,0.32}
\definecolor{lightgray}{rgb}{0.70,0.70,0.70}

%% file: commands.tex
\def\OCP{optimal control problem}

\newcommand*{\mtext}[1]{\text{\normalfont #1}} 

\newcommand*{\x}{\ensuremath{x}}%
\newcommand*{\xc}[1]{\ensuremath{\x(#1)}}%
\newcommand*{\xcdot}[1]{\ensuremath{\dot{\x}(#1)}}%
\newcommand*{\xref}{\ensuremath{\x_\text{ref}}}%

\renewcommand*{\u}{\ensuremath{u}}%
\newcommand*{\uc}[1]{\ensuremath{\u(#1)}}%

\newcommand*{\xs}{\ensuremath{\x_{\mtext{s}}}}%
\newcommand*{\ts}{\ensuremath{t_{\mtext{s}}}}%
\newcommand*{\xsolfun}{\varphi}

\newcommand*{\f}{\ensuremath{f}}

\newcommand*{\xcel}[2]{\ensuremath{\x_{#2} (#1)}}%
\newcommand*{\xu}{\ensuremath{\x_u}}
\newcommand*{\xuc}[1]{\ensuremath{\x_u(#1)}}%
\newcommand*{\xucstar}[1]{\ensuremath{\x^*_u(#1)}}%
\newcommand*{\xud}[1]{\ensuremath{\x_{#1}}}%
\newcommand*{\xudscalar}[1]{\ensuremath{x_{#1}}}%
\newcommand*{\xucdot}[1]{\ensuremath{\dot{\x}_u(#1)}}%

\newcommand*{\uuc}[1]{\ensuremath{\u(#1)}}%
\newcommand*{\uud}[1]{\ensuremath{\u_{#1}}}%
\newcommand*{\uudscalar}[1]{\ensuremath{u_{#1}}}%

\newcommand*{\uustar}{\ensuremath{\u^*}}%
\newcommand*{\uucstar}[1]{\ensuremath{\u^*(#1)}}%
\newcommand*{\uucstarbig}[1]{\ensuremath{\u^*\big(#1\big)}}%

\newcommand*{\vardt}{\ensuremath{\Delta t}}

\newcommand*{\vardtmax}{\ensuremath{\vardt_\mtext{max}}}

\newcommand*{\vardtmin}{\ensuremath{\vardt_\mtext{min}}}

\newcommand*{\Nmin}{\ensuremath{N_\mtext{min}}}%
\newcommand*{\xmu}{\ensuremath{\x_\mu}}%
\newcommand*{\xmuc}[1]{\ensuremath{\xmu{(#1)}}}%
\newcommand*{\rockets}{\ensuremath{s}}
\newcommand*{\rocketv}{\ensuremath{v}}
\newcommand*{\rocketm}{\ensuremath{m}}

\newcommand*{\xf}{\ensuremath{\x_\mtext{f}}}%

\newcommand*{\tmu}{\ensuremath{t_\mu}}%
\newcommand*{\tmud}[1]{\ensuremath{t_{\mu,#1}}}%
\newcommand*{\tf}{\ensuremath{t_\mtext{f}}}
\newcommand*{\tfstar}{\ensuremath{t^*_\mtext{f}}}

\newcommand*{\mulaw}{\ensuremath{\mu}}%

\newcommand*{\nullvec}{\ensuremath{0}}%

\newcommand*{\realposzero}{\ensuremath{\mathbb{R}^+_0}}%
\newcommand*{\naturalzero}{\ensuremath{\mathbb{N}_0}}%
\newcommand*{\naturalpos}{\ensuremath{\mathbb{N}}}

\newcommand*{\xset}{\ensuremath{\mathcal{X}}}%
\newcommand*{\rxset}{\ensuremath{\mathbb{X}}}%
\newcommand*{\uset}{\ensuremath{\mathcal{U}}}%
\newcommand*{\ruset}{\ensuremath{\mathbb{U}}}%

\newcommand*{\xfset}{\ensuremath{\mathbb{X}_\mtext{f}}}%

\newcommand*{\optimparamglobal}[1]{\ensuremath{\mathcal{P}_\mtext{global}}}

\newcommand*{\tcpumed}{\ensuremath{\Delta t_\mtext{cpu}}}

\newcommand*{\ierr}[1]{\ensuremath{e_{\hat{x}}(#1)}}

\DeclareMathOperator{\nablaop}{\nabla}
\newcommand*{\gradwrt}[2]{\ensuremath{\nablaop_{\kern -0.2em #2} #1}} 

\newcommand*{\hessianwrt}[2]{\ensuremath{\nablaop^2_{\kern -0.2em #2} #1}}

\newcommand*{\KLfun}{\ensuremath{\mathscr{K\kern -0.3em L}}}

\newcommand{\dt}{\text{d}t}
\newcommand{\dtau}{\text{d}\tau}

\newcommand{\transpose}{\intercal}

\newcommand{\fundef}[3]{#1\,{:}\,#2\,{\mapsto}\,#3}
\newcommand{\mdef}{\vcentcolon=}

\DeclarePairedDelimiter\abs{\lvert}{\rvert}%
\DeclarePairedDelimiter\norm{\lVert}{\rVert}%

\makeatletter
\let\oldabs\abs
\def\abs{\@ifstar{\oldabs}{\oldabs*}}
\let\oldnorm\norm
\def\norm{\@ifstar{\oldnorm}{\oldnorm*}}
\makeatother

\newcommand{\refsec}[1]{Section~\ref{#1}}

\newcommand{\reffigc}[1]{Figure~\ref{#1}}

\newcommand{\refsecc}[1]{Section~\ref{#1}}

%% file: theorems.tex
\newtheorem{rem}{Remark}

%% file: tikzdef.tex
\usepgfplotslibrary{patchplots}
\usetikzlibrary{positioning,automata,through,math,calc,plotmarks,shapes,arrows,arrows.meta,quotes,decorations.markings,shapes.misc,shapes,backgrounds,patterns,angles,babel,intersections,decorations,spy,decorations.pathreplacing, shadows,shadings} 
\pgfplotsset{compat=newest}
\pgfplotsset{plot coordinates/math parser=false}

\pgfplotsset{/pgf/number format/.cd,1000 sep={}} 

\newlength\figureheight
\newlength\figurewidth

\tikzstyle{every picture}+=[remember picture]
\tikzstyle{arrow} = [->,>=stealth']
\tikzstyle{arrowreverse} = [<-,>=stealth']
\tikzstyle{arrowbidir} = [<->,>=stealth']

\tikzset{draw/.append style={line width= 0.3mm}}

\tikzset{thin/.style ={line width= 0.2mm}}
\tikzset{thick/.style ={line width= 0.4mm}}
\tikzset{very thick/.style ={line width= 0.5mm}}

\pgfplotsset{tick style={black}} 

\pgfplotsset{every axis plot post/.append style={line join=round}}

\tikzset{every picture/.style={font issue=\small},
	font issue/.style={execute at begin picture={#1\selectfont}}
}
\tikzset{fontscale/.style = {font=\small}
}

\pgfplotsset{every axis/.append style={
		axis lines = left 
	}
}

\newlength\BlockSep
\newlength\BlockHeight
\newlength\BlockWidth
\newlength\sepsepsep
\tikzset{%
	partially dashed/.style={
		decoration={show path construction, 
			lineto code={
				\draw[] (\tikzinputsegmentfirst) --($(\tikzinputsegmentfirst)!#1!(\tikzinputsegmentlast)$);,
				\draw[dashed, dash phase=3pt,-latex] ($(\tikzinputsegmentfirst)!#1!(\tikzinputsegmentlast)$)--(\tikzinputsegmentlast);,
			}
		},
		decorate
	},
}

\makeatletter
\pgfplotsset{ 
	every axis x label/.append style={
		alias=current axis xlabel,
	},
	legend pos/outer south/.style={
		/pgfplots/legend style={
			at={%
				(%
				\@ifundefined{pgf@sh@ns@current axis xlabel}%
				{xticklabel cs:0.5}%
				{current axis xlabel.south}%
				)%
			},
			anchor=north,
			legend columns=3,
			font=\small,
			fill=none,
			/tikz/every even column/.append style={column sep=10pt, font=\small} 
		}
	},
	legend pos/outer north/.style={
		/pgfplots/legend style={
			at={(\figurewidth/2,\figureheight+0.2cm)},
			draw=none, 
			anchor=south,
			legend columns=3,
			font=\scriptsize,
			fill=none,
			/tikz/every even column/.append style={column sep=10pt, font=\scriptsize} 
		}
	},
	legend pos/outer north2/.style={
		/pgfplots/legend style={
			at={(\figurewidth/2,\figureheight+0.5cm)},
			draw=none, 
			anchor=south,
			legend columns=3,
			font=\scriptsize,
			fill=none,
			/tikz/every even column/.append style={column sep=5pt, font=\scriptsize} 
		}
	}
}
\makeatother

\newlength\customshift 

%% file: introduction.tex
\section{Introduction}

Time-optimal control plays an important role in many industrial areas, especially in improving the productivity of automation solutions.
In contrast to standard optimal control problems, time-optimal control demands for a variable final time in the problem formulation subject to optimization.
Therefore different or extended techniques are required to solve these problems.
Whereas simple problems can still be solved analytically with the maximum principle, generic problems for nonlinear systems are usually solved numerically with direct or indirect methods. 
An established technique for both direct and indirect methods is the time transformation approach~\cite{quintana1973_joc}.
Hereby, the time profile for state and control input trajectories is set to the unit interval, allowing cost and constraint terms to be defined similar to a fixed grid.
In addition, the system dynamics are scaled by a dedicated optimization parameter representing the true final time. 
Related second-order sufficient conditions are provided in~\cite{maurer2002_joco}.  

An established extension to the time transformation is the so-called control parameterization enhancing transform~\cite{teo1999_joams}.
The optimal control is parameterized w.r.t. a temporal grid.
This grid and the corresponding switching times are mapped onto a uniformly spaced grid in a new fixed time scale but with an individual optimization parameter and scaled system dynamics for each grid partition.
Applications to different system classes are provided in~\cite{rehbock1999_chapter,li2006_mcm}. 
Related work in~\cite{vossen2010_jota} presents a direct method with optimality conditions as well as first- and second-order variational derivatives of the state trajectory w.r.t. the switching times. 
An iterative indirect method is proposed in~\cite{kashiri2011_acs} with a dedicated initialization strategy to account for the difficulties in initializing the problem, which is generally a known problem for indirect methods.
An approximate time-optimal control in the arc time space for single-input nonlinear systems is presented in~\cite{kaya2003_jota}.
A well-known application for time-optimal control is the lap time minimization of racing cars, for which a method is presented e.g. in~\cite{kelly2010_vsd}.  

More recent approaches realize time-optimal feedback control in the framework of predictive control.
General analyses and theoretical results for time-optimal feedback control of nonlinear discrete time systems are presented in~\cite{pin2014_tac,sutherland2019_cdc}.
For continuous-time systems, direct methods for the underlying optimal control problems are usually preferred to indirect methods based on larger convergence regions and their suitability for real-time optimization.
For point-to-point transitions, a common approach is to rely on time transformation for the underlying direct optimal control problem that needs to be solved in every closed-loop step.
E.g.,~\cite{zhao2004_asce} applies multiple shooting as direct method in combination with the time transformation and uses a hybrid cost function considering both minimum-time and quadratic form objectives.
Dedicated methods for time-optimal predictive control are presented in~\cite{vandenbroeck2011_mechatronics} and~\cite{verschueren2017_cdc}.

Previous work proposes minimum-time optimal control and predictive control approaches based on direct methods and variable discretization~\cite{roesmann2019_phd},
in particular in~\cite{roesmann2015_ecc} with collocation via finite differences, in \cite{roesmann2020_stability} with multiple shooting and in~\cite{roesmann2017_cdc} with a non-uniform shooting grid for bang-singular-bang systems.

This paper addresses the time-optimal control problem formulation based on variable discretization and direct collocation via quadrature.
The basic idea of direct collocation is to approximate the dynamics, cost and constraint functions by a set of basis functions which are defined w.r.t. grid/knot points~\cite{betts2010_book}.
Established candidates for basis functions are quadrature rules often resulting in piecewise linear, quadratic or cubic splines for the states and control trajectories~\cite{kelly2017_siam}.
Direct collocation usually requires a larger number of optimization parameters compared to multiple shooting, but is well suited for stiff systems and achieves higher accuracies, especially for tasks requiring more complex control trajectories.
Note, orthogonal collocation and pseudospectral methods are established specializations in the literature~\cite{ross2012_aric}.
This work accounts for a potential applicability for closed-loop predictive control and real-time optimization.
Therefore, we choose lower-order polynomials such as in Hermite-Simpson collocation~\cite{kelly2017_siam} as representative. 
A special focus of this paper is the comparison of the different control parameterizations.
To our best knowledge, these results are not yet available in the literature and are of high practical relevance.

The next section introduces the formal description of the time-optimal control problem.
\refsecc{sec:collocation} proposes the direct collocation formulations with variable discretization.
The evaluation and analysis is conducted in \refsecc{sec:example} and \refsec{sec:summary} concludes the work.

%% file: theory.tex
\section{Preliminaries and Problem Setup}
\label{sec:prelim}

\subsection{Dynamic System}
\label{sec:dyn_sys}
We consider continuous-time, nonlinear, time-invariant systems with state trajectory $\fundef{\x}{\mathbb{R}}{\xset}$ and control trajectory $\fundef{\u}{\mathbb{R}}{\uset}$:
\begin{equation}
\xcdot{t} = \f\big(\xc{t},\uc{t}\big). 
\label{eq:system_dynamics_cont}
\end{equation}
Throughout this paper, the state space $\xset$ is defined as $\xset \mdef \mathbb{R}^p$ with state vector dimension $p\in\naturalpos$. 
The control space $\uset$ is given by $\uset \mdef \mathbb{R}^q$ with control vector dimension $q\in\naturalpos$. 
Function $\fundef{\f}{\xset \times \uset}{\xset}$ defines a nonlinear mapping of the state and control trajectory, $\xc{t}$ and $\uc{t}$ respectively, to the state velocity $\xcdot{t}$ embedded in $\xset$.
System~\eqref{eq:system_dynamics_cont} is further subject to state and input constraint sets, i.e. $\xc{t} \in \rxset \subseteq \xset$ and $\uc{t} \in \ruset \subset \uset$, respectively.
The solution to~\eqref{eq:system_dynamics_cont} contained in an open time interval $I \subseteq \mathbb{R}$ with initial value $\xc{\ts}=\xs$, $\ts \in I$, $\xs \in \xset$ and $t \in I$ is defined by 
\begin{equation}
\xsolfun\big(t, \xs, \uc{t}\big) = \xs + \int_{\ts=0}^{t} \f\big(\xc{\tau},\uc{\tau}\big) \dtau.
\label{eq:ivp:x}
\end{equation}
Without loss of generality, initial time $\ts$ is fixed to $\ts=0$ as~\eqref{eq:system_dynamics_cont} is time-invariant.
Carath\'{e}odory's existence theorem addresses conditions for the existence and uniqueness of the solution.
In the following, we assume that the vector field $\f$ is continuous and Lipschitz in its first argument.
Furthermore, the control $\uc{t}$ is supposed to be locally Lebesgue integrable for $t \in I$, i.e. $u \in L^\infty(I, \uset)$.

\subsection{Optimal Control Problem}
\label{sec:ocp}

The optimal control task comprises the transition from an initial state $\xuc{t_0}=\xs$ at time $t_0\in I$ to some terminal set $\xfset \subseteq \rxset$, i.e. $\xu(\tf) \in \xfset$, in minimum time $\tf \in I$. 
The associated continuous-time time-optimal control problem with state constraints $\xuc{t} \in \rxset$ and control constraints $\uuc{t} \in \ruset$ is given as follows:
\begin{gather}
    \tfstar(\xs) = \underset{\uuc{t}, \tf}{\min}\ \tf
	\label{eq:cont_time_to_ocp} \\
	\hspace{-7.3cm}\text{subject to} \nonumber \\ 
	\begin{align*}
		&\xuc{t_0=0} = \xs, \quad \xuc{t} \in \rxset, \quad \uuc{t} \in \ruset, \quad \xuc{\tf} \in \xfset, \\ 
		&\xucdot{t} = \f \big(\xuc{t},\uuc{t}\big).
	\end{align*}
\end{gather}
We denote the optimal control and state trajectories by $\uucstar{t, \xs}$ and $\xucstar{t, \xs}$ respectively while emphasizing their relation to initial state $\xs$.
Accordingly, the minimum transition time is given by $\tfstar(\xs)$.
Note that $\xucstar{t, \xs} = \xsolfun\big(t, \xs,\uucstar{t, \xs}\big)$ holds according to~\eqref{eq:ivp:x} with $t\in [0,\tf]$.
A control trajectory $u(t)$ and the corresponding state trajectory $\xuc{t}$ are called \textit{admissible for $\xs$ up to time~$\tf$}
if $u(t) \in \ruset$, $\xuc{t} \in \rxset$ and $\xuc{\tf} \in \xfset$ hold for $t\in [0,\tf]$. 
The \OCP{}~\eqref{eq:cont_time_to_ocp} is referred to as \textit{feasible} if $\uucstar{t, \xs}$ and $\xucstar{t, \xs}$ are admissible from $\xs$ up to time $\tfstar(\xs,N)$.

\section{Direct Collocation}
\label{sec:collocation}

Direct collocation discretizes both the control and state trajectories in~\eqref{eq:cont_time_to_ocp} according to a specified grid in order to transform it to a nonlinear program which can be solved by standard parameter optimization techniques.
Let $0 = t_0 \leq t_1 \leq \dotsc \leq t_k \leq \dotsc \leq t_N = \tf$ with $t_k,\tf \in I$, $k=0,1,\dotsc,N$ and $N\in\naturalpos$ define the discretization grid. 
Partitions are further restricted to $t_{k+1}-t_k = \vardt$ to ensure uniformity with interval lengths $\vardt \in \realposzero$.
Accordingly, $t_k = k\vardt$ refers to individual grid points. 
We further denote the control values at time instances $t_k$ as $\uuc{t_k} \mdef \uud{k}$ and states as $\xuc{t_k} \mdef \xud{k}$ respectively.
For every grid partition, collocation via numerical quadrature approximates the following integral form of~\eqref{eq:system_dynamics_cont}~\cite{kelly2017_siam}:
\begin{equation}
\xud{k+1} - \xud{k}   = \int_{t_k}^{t_k + \vardt} \f\big(\xuc{t}, \uuc{t}\big)\,\dt. 
\label{eq:hs_integral_eq}
\end{equation}

\begin{figure}[tb]
	\centering
	
	\tikzstyle{style1}=[black,solid,line width=1.25pt]
	\tikzstyle{style2}=[darkgray,solid,line width=1.25pt] 
	\tikzstyle{style3}=[black,dashed,line width=1.25pt]
	\tikzstyle{style4}=[lightgray,solid,line width=1.25pt] 
	\tikzstyle{style4b}=[black,dotted,line width=1.25pt] 
	\tikzstyle{style5}=[darkgray,dashed,line width=1.25pt] 
	\tikzstyle{style6}=[lightgray,dashed,line width=1.25pt] 
	\tikzstyle{style7}=[darkgray, dash pattern={on 7pt off 2pt on 1pt off 3pt},line width=1.25pt]
	
		\begin{subfigure}[b]{.48\columnwidth}
		\centering
		\begin{tikzpicture}[font=\footnotesize]
		\setlength{\figurewidth}{1.2\columnwidth}
		\setlength{\figureheight}{4.75cm}
		\begin{axis}[
		axis lines=center,
		xtick=\empty, ytick=\empty,
		x axis line style={draw=none}, 
		y axis line style={draw=none}, 
		xmin=-0.1,xmax=1.7,
		ymin=-0.8,ymax=1.0,
		xlabel=$\color{darkgray}t$,
		ylabel=$\color{darkgray}x(t)$,
		xtick={1.5},
		xticklabel={$\vardt$},
		width=\figurewidth,
		height=\figureheight,
		anchor=center,
		clip=false 
		]
		\draw [/pgfplots/every inner x axis line, draw=darkgray, line cap=rect] (axis cs:0,0) -- (axis cs:\pgfkeysvalueof{/pgfplots/xmax}, 0);
		\draw [/pgfplots/every inner y axis line, draw=darkgray, line cap=rect] (axis cs:0,0) -- (axis cs:0,\pgfkeysvalueof{/pgfplots/ymax});
		
		\pgfmathsetmacro{\dt}{1.5}%

		\coordinate (x1) at (0, 0.5);
		\coordinate (x15) at (0.75, 0);
		\coordinate (x2) at (1.5, -0.5);
		
		\draw[style1] (x1) to[in=120,out=10] (x15);
		\draw[style1] (x15) to[in=190,out=300] (x2);
		
		\draw[black, fill=black] (x1) circle (0.09cm);
		\draw[fill=white] (x15) circle (0.07cm);
		\draw[fill=black] (x2) circle (0.09cm);

		\node[below left=0.4cm and 0cm, align=center] at (x15) (mplabel) {Midpoint\\$\xudscalar{k+0.5}$};
		\draw[-latex, shorten >= 1mm] (mplabel) -- (x15);
		
		\node[below right] at (x1) {$\xudscalar{k}$};
		\node[above right=0.0cm and -0.1cm] at (x2) {$\xudscalar{k+1}$};
		
		\draw[darkgray, rotate around={120:(x15)}] (0.45,0) -- (1.05,0);			
		\draw[darkgray, rotate around={10:(x1)}] (x1) -- (0.3,0.5);
		\draw[darkgray, rotate around={10:(x2)}] (x2) -- (1.2,-0.5);
		\end{axis}
		\end{tikzpicture}
		\caption{Cubic Hermite state spline}
		\label{fig:hs:state_spline}
	\end{subfigure}
	\begin{subfigure}[b]{.48\columnwidth}
		\centering
		\begin{tikzpicture}[font=\footnotesize]
		\setlength{\figurewidth}{1.2\columnwidth}
		\setlength{\figureheight}{4.75cm}
		
		\begin{axis}[
		axis lines=center,
		xtick=\empty, ytick=\empty,
		x axis line style={draw=none}, 
		y axis line style={draw=none}, 
		xmin=-0.1,xmax=1.7,
		ymin=-0.8,ymax=1.0,
		xlabel=$\color{darkgray}t$,
		ylabel=$\color{darkgray}u(t)$,
		xtick={1.5},
		xticklabel={$\vardt$},
		width=\figurewidth,
		height=\figureheight,
		anchor=center,
		clip=false, 
		legend columns = 2,
		legend pos = outer north,
		legend style={draw=none,fill=none,legend cell align=left, column sep =4pt, font=\scriptsize,at={(\figurewidth/2-0.6cm,\figureheight-1.5cm)}},
		]
		\draw [/pgfplots/every inner x axis line, draw=darkgray, line cap=rect] (axis cs:0,0) -- (axis cs:\pgfkeysvalueof{/pgfplots/xmax}, 0);
		\draw [/pgfplots/every inner y axis line, draw=darkgray, line cap=rect] (axis cs:0,0) -- (axis cs:0,\pgfkeysvalueof{/pgfplots/ymax});
		
		\pgfmathsetmacro{\dt}{1.5}%
		
		\pgfmathsetmacro{\tone}{0}%
		\pgfmathsetmacro{\uone}{0.5}%
		\pgfmathsetmacro{\uonemid}{-0.5}%
		\pgfmathsetmacro{\usecond}{0.25}%
		
		\pgfmathsetmacro{\tonemid}{\tone+0.5*\dt}%
		\pgfmathsetmacro{\tsecond}{\tone+\dt}%
		\pgfmathsetmacro{\betaone}{-1/\dt*(3*\uone - 4*\uonemid + \usecond)}%
		\pgfmathsetmacro{\betatwo}{2/(\dt^2)*(\uone - 2*\uonemid + \usecond)}%
		\addplot[style1, restrict x to domain=0:\dt, samples=300, forget plot] {\uone + \betaone*(x-\tone) + \betatwo*(x-\tone)^2};
		\edef\ta{\tone}
		\edef\ua{\uone}
		\edef\tb{\tonemid}
		\edef\ub{\uonemid}
		\edef\tc{\tsecond}
		\edef\uc{\usecond}
		
		\coordinate (u1) at (\ta, \ua);
		\coordinate (u15) at (\tb, \ub);
		\coordinate (u2) at (\tc, \uc);
		
		\draw[style3] (u1) -- (u15);
		\draw[style3] (u15) -- (u2);
		
		\draw[style4b] (u1) -- (u2);
		
		\draw[style7] (u1) -- (u1 -| u2);
		
		\draw[black, fill=black] (u1) circle (0.09cm);
		\draw[fill=white] (u15) circle (0.07cm);
		\draw[fill=black] (u2) circle (0.09cm);
		
		\node[below] at (u15) {Midpoint $\uudscalar{k+0.5}$};
		
		\node[above right] at (u1) {$\uudscalar{k}$};
		\node[above right] at (u2) {$\uudscalar{k+1}$};

		\addlegendimage{style1};
		\addlegendentry{Quadratic};
		\addlegendimage{style3};
		\addlegendentry{Linear};
		\addlegendimage{style4b};
		\addlegendentry{Mean};
		\addlegendimage{style7};
		\addlegendentry{Constant};
		
		\end{axis}
		\end{tikzpicture}
		\caption{Control representations}
		\label{fig:hs:control_spline}
	\end{subfigure}%
	\caption{State and control representations of Hermite-Simpson collocation.}
	\label{fig:hs_splines}
\end{figure}
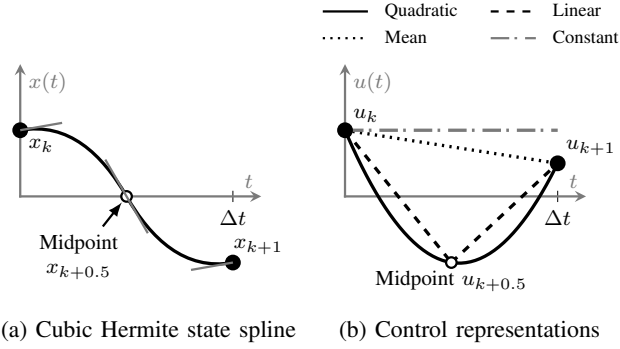

In Hermite-Simpson collocation, the Simpson quadrature rule approximates the integrand in~\eqref{eq:hs_integral_eq} for $t \in [t_k, t_{k}+\vardt]$ by a quadratic polynomial~\cite{kelly2017_siam}:
\begin{align}
&\int_{t_k}^{t_k + \vardt}  \f \big(\xuc{t}, \uuc{t} \big)\,\dt \label{eq:uniform:hs:phi} \\ 
&\approx \underbrace{\frac{1}{6} \vardt \big( \f(\xud{k}, \uud{k}) + 4 \f(\hat{x}_{k+0.5}, \hat{u}_{k+0.5} ) + \f(\xud{k+1}, \uud{k+1}) \big)}_{=\xi(\xud{k}, \uud{k}, \hat{x}_{k+0.5}, \hat{u}_{k+0.5}, \xud{k+1}, \uud{k+1})}. \nonumber
\end{align}
State and controls at midpoints $t_{k+0.5} \mdef 0.5 ( t_{k+1} + t_k )$ of the $k$-th grid partition are denoted as $\hat{x}_{k+0.5}  \mdef \xuc{t_{k+0.5}}$ and $\hat{u}_{k+0.5} \mdef \uuc{t_{k+0.5}}$ respectively.
States $\xud{k}$ and $\xud{k+1}$ as well as controls $\uud{k}$ and $\uud{k+1}$ coincide with grid points $t_k$ but $\hat{x}_{k+0.5}$ is not known in advance.
Fortunately, $\hat{x}_{k+0.5}$ is computed from a quadratic interpolant by evaluating the states and function values at grid points $k$ and $k+1$:
\begin{equation}
\hat{x}_{k+0.5} \mdef \frac{1}{2} ( \xud{k} + \xud{k+1} ) + \frac{\vardt}{8} \big(  \f(\xud{k}, \uud{k}) - \f(\xud{k+1}, \uud{k+1}) \big).
\label{eq:uniform:hs:xmid}
\end{equation}
The actual derivation is provided in~\cite{kelly2017_siam}. 
Equation~\eqref{eq:uniform:hs:xmid} becomes a separate equality constraint to the nonlinear program which is referred to as uncompressed form.
Otherwise $\xud{k+0.5}$ is replaced directly in~\eqref{eq:uniform:hs:phi}. 
The latter is denoted as compressed form.
Choosing the compressed or uncompressed form has no influence on the actual solution, but on the number of parameters to be optimized and the problem structure. 
Approximating the dynamics by a quadratic spline results in a cubic Hermite spline for the optimal state trajectory as shown in \reffigc{fig:hs:state_spline}.   

The nonlinear program obtained from \eqref{eq:cont_time_to_ocp} and direct transcription is generally defined as follows:
\begin{gather}
	\tfstar(\xs,N) = \underset{ \substack{ \uud{k} \forall k \in \mathcal{I}_\mtext{u}, \\ \xud{k} \forall k \in \mathcal{I}_\mtext{x}, \\ \vardt} }{\min}\ N\vardt
	\label{eq:hs:nlp} \\
	\hspace{-7.3cm}\text{subject to} \nonumber \\ 
	\begin{align*}
		&\xud{0} = \xs, \quad \xud{N} \in \xfset,  \quad \vardtmin \leq \vardt \leq \vardtmax,\\ 
		&\xud{k} \in \rxset \mtext{ for all  } k = 0.5,1,\dotsc,N,\\
		&\uud{k} \in \ruset \mtext{ for all } k \in \mathcal{I}_\mtext{u},\\
		& \phi ( \mathbf{\x}_k, \uud{k}, \hat{u}_{k+0.5}, \hat{u}_{k+1}, \vardt ) = \nullvec \mtext{ for all } k = 0, 1, \dotsc, N-1.
	\end{align*}
\end{gather}
Parameter $\tf$ in \eqref{eq:cont_time_to_ocp} is hereby substituted by the local time interval $\vardt$ as $\tf = N\vardt$ holds.
The bounds $\vardtmin,\vardtmax \in \realposzero$ with $\vardtmax \geq \vardtmin$ are introduced for technical reasons and their purpose is described later.
$\mathcal{I}_\mtext{x}$ and $\in \mathcal{I}_\mtext{u}$ denote the indexes sets for the state and control parameters subject to optimization.
Controls $\hat{u}_k$ and $\hat{u}_{k+1}$ are placeholders and are defined later.
$ \phi ( \cdot )$ is the collocation constraint and for the compressed form it is:
\begin{align}
\phi ( \cdot) = \xi(\xud{k}, \uud{k}, \hat{x}_{k+0.5}, \hat{u}_{k+0.5}, \xud{k+1}, \hat{u}_{k+1})
\end{align} 
with $\mathbf{\x}_k = [\xud{k}, \xud{k+1}]$ and $\mathcal{I}_\mtext{x} = \{0,1,\dotsc, N\}$.
Note that in this formulation, midpoints $\xud{k+0.5}\mdef\hat{x}_{k+0.5}$ computed by~\eqref{eq:uniform:hs:xmid} are still subject to constraint evaluation. 

In contrast, the collocation constraint for the uncompressed form is given by:
\begin{align}
\thinmuskip 2mu
\medmuskip 1mu
\phi ( \cdot) = \begin{pmatrix} \medmath{\xi(\xud{k}, \uud{k}, {x}_{k+0.5}, \hat{u}_{k+0.5}, \xud{k+1}, \hat{u}_{k+1})} \\ \medmath{x_{k+0.5} - \frac{1}{2} ( \xud{k} + \xud{k+1} ) - \frac{\vardt}{8} \big(  \f(\xud{k}, \uud{k}) - \f(\xud{k+1}, \hat{u}_{k+1}) \big)} \end{pmatrix} \nonumber
\end{align} 
\thinmuskip 3mu
\medmuskip 4mu
with $\mathbf{\x}_k = [\xud{k}, \xud{k+0.5}, \xud{k+1}]$ and $\mathcal{I}_\mtext{x} = \{0,0.5,1,\dotsc, N\}$. 

Note, the previous derivation does not consider any particular choice for the control parameterization between grid and midpoints.
Hence, we consider the possible cases that correspond to different degrees of freedom in control.
\reffigc{fig:hs:control_spline} illustrates theses cases and their associated optimization parameters.
Only the quadratic and linear spline consider $\uud{k+0.5}$ as explicit optimization parameter.

\subsection{Quadratic and Linear Control Spline}

The full degrees of freedom in control are obtained by including the midpoint control $\u_{k+0.5}$ as separate optimization parameter.
The related index set is $\mathcal{I}_{\mtext{u}} = \{ 0, 0.5, 1 \dotsc, N\}$ and the control placeholders are substituted by the actual optimization parameters, i.e. $\hat{\u}_{k+0.5} \mdef \u_{k+0.5}$ and $\hat{\u}_{k+1} \mdef \u_{k+1}$.
Reconstructing the continuous-time control trajectory $u(t)$ for $t \in [0, \tf]$ after solving~\eqref{eq:hs:nlp} is performed with one of the two following interpolation schemes:

The \textit{quadratic control spline} $\uuc{t}$ follows from a quadratic polynomial for each grid partition $t \in [t_k, t_{k+1}]$:
\begin{align}
	\uuc{t} &\mdef \uud{k} + \beta_1 (t-t_k) + \beta_2 (t-t_k)^2, \label{eq:uniform:hs:u} \\
	\beta_1 &= -\frac{1}{\vardt}( 3 \uud{k} - 4 \uud{k+0.5} + \uud{k+1} ), \nonumber \\
	\beta_2 &= \frac{2}{\vardt^2}(  \uud{k} - 2 \uud{k+0.5} + \uud{k+1}   ). \nonumber
\end{align}

The \textit{linear control spline} is defined by linear segments between $\uud{k}$ and $\uud{k+0.5}$ as well as $\uud{k+0.5}$ and~$\uud{k+1}$:
\begin{align}
&\uuc{t} = \\
&\begin{cases}
 u_k\! +\! (t\!-\!t_k)(u_{k+0.5}\!-\!u_k) & \hspace{-0.2cm}\mtext{for } t\in[t_k,t_{k+0.5}] \\
 u_{k+0.5}\! +\! (t\!-\!t_{k+0.5})(u_{k+1}\!-\!u_{k+0.5}) & \hspace{-0.2cm}\mtext{for } t\in(t_{k+0.5},t_{k+1}] 
\end{cases} \nonumber
\end{align}

Since the number of parameters is identical for both control parameterizations, they are represented by the very same nonlinear program.

\subsection{Linear Mean Control Spline}

Omitting $\uud{k+0.5}$ as additional optimization parameter and substituting the midpoint control placeholder by $\hat{u}_{k+0.5} = 0.5\big(\uud{k} + \uud{k+1}\big)$ results in the \textit{(linear) mean control spline}:
\begin{equation}
u(t) = u_k + (t-t_k)(u_{k+1}-u_k) \mtext{ for } t \in[t_k, t_{k+1}]. 
\end{equation}
The index set for control parameters is $\mathcal{I}_{\mtext{u}} = \{ 0,1,\dotsc,N\}$ and placeholder $\hat{\u}_{k+1}$ is set to $\hat{\u}_{k+1} \mdef \u_{k+1}$.

\subsection{Piecewise Constant Control}

The least degrees of freedom in control is obtained by a \textit{piecewise constant control trajectory}
that omits both the midpoints $\uud{k+0.5}$ and the final control~$\uud{N}$:
\begin{equation}
u(t) = u_k \mtext{ for } t \in[t_k, t_{k+1}]. 
\end{equation}
The corresponding index set for control parameters is similar as for the linear mean control spline, i.e. $\mathcal{I}_{\mtext{u}} = \{ 0,1,\dotsc,\allowbreak N-1\}$.
For this representation, the control placeholders are $\hat{u}_{k+1} \mdef \hat{u}_{k+0.5} \mdef u_{k}$.

\subsection{Feasibility and Optimality}
\label{sec:feas_opt}

The nonlinear program~\eqref{eq:hs:nlp} approximate the continuous-time time-optimal control problem \eqref{eq:cont_time_to_ocp} with $N$ grid partitions of length $\vardt$ each.
Hereby, the grid size $N$ is crucial for the accuracy and feasibility of the solutions.
The upper time bound $\vardtmax > 0$ makes it possible to bind a worst-case accuracy to the feasibility property.
Note that $\vardtmax$ might be determined by analyzing the system dynamics a-priori.
To ensure that a solution to the optimal control problem exists, assume that there is an $N>0$ for which~\eqref{eq:hs:nlp} is feasible.
In addition, necessary and sufficient optimality conditions for general nonlinear programs apply~\cite{nocedal2006_book}.
Any practical implementation replaces compact and convex constraint sets $\rxset,\ruset$ and $\xfset$ by algebraic equality and inequality constraint functions.

\begin{rem}
	NLP~\eqref{eq:hs:nlp} is derived w.r.t. a global optimization parameter $\vardt$ for each time interval $[t_k,t_{k+1}]$ (global uniform grid approach).
	Another formulation is obtained by replacing $\vardt$ by individual time parameters $\vardt_k$ for each time interval (local uniform grid approach). 
	Uniformity must be enforced by adding additional constraints $\vardt_k = \vardt_{k+1}$ for $k=0,1,\dotsc,N-1$ to~\eqref{eq:hs:nlp}.
	Although the optimal solution is identical, the structure of the optimization problem differs slightly. Benchmarks~\cite{roesmann2019_phd} show comparable commutation times for small to medium-sized control tasks.
\end{rem}

\subsection{Feedback Control}

Even feedback control is not in the scope of this paper, we would like to highlight that the proposed approach can be seamlessly integrated into a shrinking-horizon predictive control scheme.
Details on the realization are provided in \cite{roesmann2020_stability} for multiple shooting but also apply to direct collocation as shown here.
In summary, nonlinear program \eqref{eq:hs:nlp} is solved only at discrete time instances
$\tmud{0} < \tmud{1} < \dotsc < \tmud{n} < \dotsc < \infty$ with \mbox{$n \in \naturalzero$} and $\tmud{n} \in \realposzero$.
The control law $\fundef{\mulaw}{\xset}{\uset}$ for closed-loop time $\tmu \in [\tmud{n}, \tmud{n+1})$ and measured or observed state feedback $\xmuc{\tmu}$ is given by:
\begin{equation}
\mulaw\big(\xmuc{\tmu}\big) \mdef \uucstarbig{\tmu-\tmud{n}, \xc{\tmud{n}}} \big\vert_{N=N_n}.
\label{eq:mulaw}
\end{equation}
Hereby, $\uucstarbig{\tmu-\tmud{n}, \xc{\tmud{n}}}$ denotes the optimal solution of $\eqref{eq:hs:nlp}$ as defined in~\refsec{sec:ocp}. The horizon length $N$ is substituted by $N=N_n$ with initial horizon length $N_0 \in \naturalpos$.
In case the very first optimal control problem is feasible, a grid adaptation scheme reduces the current horizon length by one $N_{n+1} = \max(N_n -1, \Nmin)$ and thus ensures forward invariance and convergence towards a small region around the target set depending on the choice of $\vardtmin$ and $\Nmin$. $\Nmin$ is a safe guard and is usually set to $\Nmin\geq p$ and in feedback control it is recommended to set $\vardtmin$ to some small positive value to avoid numerical ill-conditioning close to the target state.
A smooth stabilization is then optionally achieved using a dual-mode control scheme.

%% file: example.tex
\section{Evaluation and Analysis}
\label{sec:example}

 \tikzstyle{style1}=[black,dashed,line width=1.25pt]
\tikzstyle{style2}=[black,solid,line width=1.25pt] 
\tikzstyle{style3}=[darkgray,solid,line width=1.25pt] 
\tikzstyle{style4}=[lightgray,solid,line width=1.25pt] 
\tikzstyle{style5}=[darkgray,dashed,line width=1.25pt] 
\tikzstyle{style6}=[lightgray,dashed,line width=1.25pt]

The evaluation is performed with two benchmark systems in simulation.
The nonlinear programs are solved with the established C++ interior point solver IPOPT~\cite{waechter2006_matprog} and HSL-MA57 as internal linear solver~\cite{hsl}.
In our implementation, the sparsity structure is exploited by computing sparse finite differences for Jacobian and Hessian matrices based on a hypergraph representation~\cite{roesmann2018_aim}. 

The first benchmark system is the Van der Pol oscillator which is commonly reported in the literature.
It is a second-order dynamic system with nonlinear damping described by $\ddot{y}(t) - \big( 1 - y(t)^2 \big) \dot{y}(t) + y(t) = u(t)$ with $\fundef{y}{\mathbb{R}}{\mathbb{R}}$.
Transforming the differential equation to a state space model~\eqref{eq:system_dynamics_cont} with state vector $\xc{t} \mdef \big(\xcel{t}{1}, \xcel{t}{2}\big)^\transpose$, $\xset = \mathbb{R}^2$ and $\uset = \mathbb{R}$ results in:
\begin{equation}
\xcdot{t}  = \big(\xcel{t}{2},\ \big( 1 - \xcel{t}{1}^2 \big) \xcel{t}{2} - \xcel{t}{1} + u(t)\big)^\intercal.
\label{eq:vdp_system}
\end{equation}

\begin{figure}[tb]
	\centering
	\includegraphics{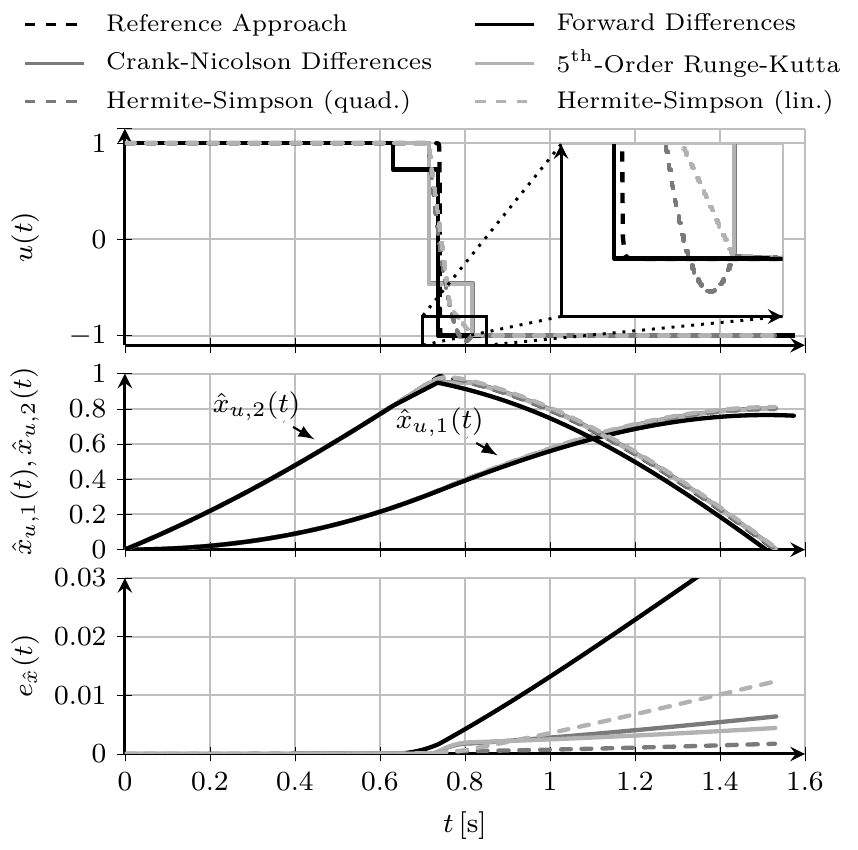}
	\caption{Optimal trajectories for the Van der Pol oscillator with $N=15$ and several direct methods. The bottom plot shows the integral error w.r.t. the reference solution.} 
	\label{fig:uniform:vdp_open_loop}
	\vspace{-\baselineskip}
\end{figure}

In the first scenario, the control task is to control the system from the origin to $\xf=(0.8,0)^\transpose$ in minimum time while constraints are set to $\vardtmin=0$, $\vardtmax=\infty$, $\rxset =\xset$ and $\ruset = \{u \in \uset \mid |u| \leq 1\}$.
The grid size is set to $N=15$.
\reffigc{fig:uniform:vdp_open_loop} shows the solutions for Hermite-Simpson collocation with quadratic and linear control parameterizations. In addition, the solutions for collocation via finite-differences (forward differences and Crank-Nicolson) and 
multiple shooting with \nth{5}-order Runge-Kutta are depicted. 
A reference time-optimal trajectory is obtained by a dedicated boundary value problem with time transformation~\cite{avvakumov2004}.
While the control trajectories $u(t)$ are obtained from the solver, the state trajectories are precisely simulated with $u(t)$ and~\eqref{eq:ivp:x}, indicated by $\hat{x}$.
The control trajectories differ at most at the control switching point at approx. \SI{0.7}{s}. The switching in control is realized over two consecutive intervals as the limited grid resolution $N=15$ cannot match the ideal switching point from the reference.
Note that the Hermite-Simpson method with quadratic control splines exceeds the control bounds as constraints are only enforced at collocation points.
To evaluate the dynamics accuracy w.r.t. the reference solution $\xref$, the integral dynamics error $\ierr{t} = \int_{0}^{t} \lVert x_\mtext{ref}(\tau) - \xsolfun\big(\tau, \xs, \uustar(\tau)\big)\rVert_2 \,\dtau$ is shown in the bottom plot.
Forward differences reveal the largest error as it is a first-order method. The smallest error is achieved with Hermite-Simpson and quadratic control splines but it violates the lower control limit as mentioned before.

\begin{figure}[tb]
	\centering
	\includegraphics{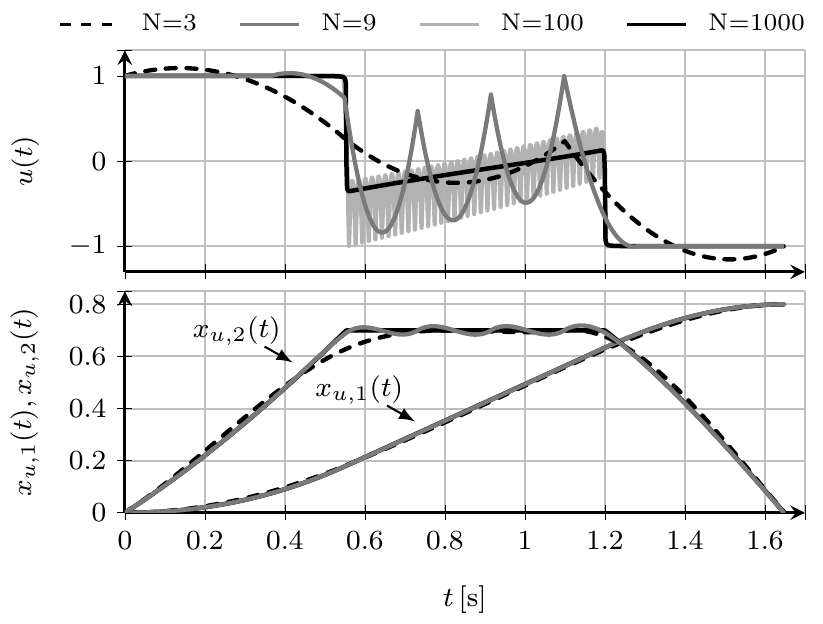}
	\caption{Solutions obtained from Hermite-Simpson collocation with quadratic control splines for the constrained Van der Pol oscillator and varying $N$.	} 
	\label{fig:uniform:vdp_hermite_simpson}
	\vspace{-0.5\baselineskip}
\end{figure}

The next scenario further investigates the impact of the additional degrees in the control parameterization.
A state constraint is added to the previous control task, i.e. $\rxset = \{x \in \xset \mid |(0,1)\x| \leq 0.7\}$.
\reffigc{fig:uniform:vdp_hermite_simpson} shows the solutions for quadratic control splines and varying grid sizes $N$.
Note that the solution for $N=3$ is already quite accurate for the state trajectory as Hermite-Simpson is an implicit method.
However, the additional degree of freedom in control at midpoints $\xud{k+0.5}$ allows the cubic state trajectory to violate state constraints between the grid and midpoints to minimize time.
With a larger grid size, the oscillations are at least visibly reduced. 
A possible remedy to avoid these oscillations is to add further constraint  evaluations to the nonlinear programs, even if this increases the calculation times.
But more suitable is the reduction of degrees of freedom in the control parameterization.

\begin{figure}[tb]
	\centering
	\includegraphics{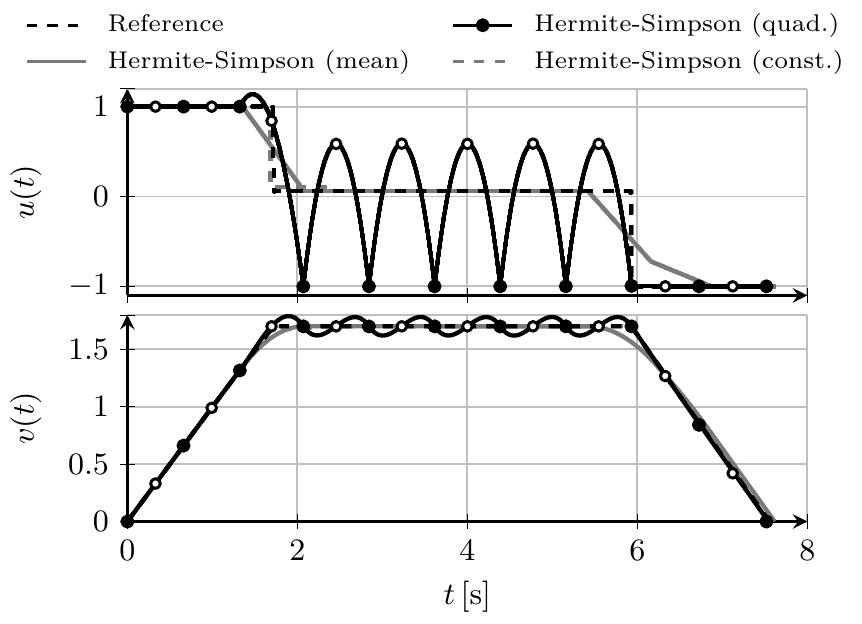}
	\caption{Rocket system solutions obtained from Hermite-Simpson collocation with different control parameterizations.}
	\label{fig:rocket_open_loop_hs}
	\vspace{-0.5\baselineskip}
\end{figure}

Consider another popular benchmark system, the free-space rocket, to demonstrate the effects.
With position $\rockets(t)\in \mathbb{R}$, velocity $\rocketv(t)\in \mathbb{R}$, mass $\rocketm(t)\in \mathbb{R}$  and  state vector $\xc{t} \mdef \big(\rockets(t), \rocketv(t), \rocketm(t)\big)^\transpose$ the dynamics are:
\begin{align}
	\begin{split}
		\xcdot{t} &=  \f\big(\xc{t}, u(t)\big) =
		\begin{pmatrix}
			\rocketv(t) \\
			\frac{u(t)-0.02\, \rocketv(t)^2}{\rocketm(t)} \\
			-0.01\, u(t)^2
		\end{pmatrix}.
	\end{split}
	\label{eq:system:rocket:ss}
\end{align} 
Constraints are set to $\rxset =\{ (\rockets,\rocketv,\rocketm)^\transpose \in \xset \mid -0.5 \leq \rocketv \leq 1.7, \rocketm \geq 0 \}$ and $\ruset = \{u \in \uset \mid |u| \leq 1\}$.
The target set is specified as $\xfset = \{ (\rockets,\rocketv,\rocketm)^\transpose \in \rxset \mid \rockets = 10, \rocketv = 0 \}$.
\reffigc{fig:rocket_open_loop_hs} shows the control trajectory and velocity profile for $N=10$ and different control parameterizations.
The linear control spline is omitted as the solution is similar to the quadratic spline, but with linearly connected  $\uud{k}, \uud{k+0.5}$ and $\uud{k+1}$.
Similar as before, the quadratic control spline oscillates and inherently violates constraints inbetween grid points. 
On the other hand, the mean and the piecewise constant control parameterizations show no oscillations.
Note that these oscillations are not limited to these two benchmark systems because they occur due to the optimal control problem definition.
\begin{figure}[tb]
	\centering
	\begin{subfigure}[b]{\columnwidth}
		\centering
		\begin{tikzpicture}[font=\footnotesize]
		\begin{axis}[hide axis,	xmin=0,	xmax=1,	ymin=0,	ymax=1,	legend columns = 2,	legend pos = outer north, legend style={draw=none,fill=none,legend cell align=left, column sep =4pt, font=\scriptsize}	]
		\addlegendimage{style1}; \addlegendentry{Uncompressed (quad.)};
		\addlegendimage{style4}; \addlegendentry{Compressed (quad.)};
		\addlegendimage{style2}; \addlegendentry{Uncompressed (const.)}
		\addlegendimage{style5}; \addlegendentry{Compressed (const.)}
		\end{axis}
		\end{tikzpicture}
	\end{subfigure}	\\
	\begin{subfigure}[b]{0.49\columnwidth}
		\centering
		\includegraphics{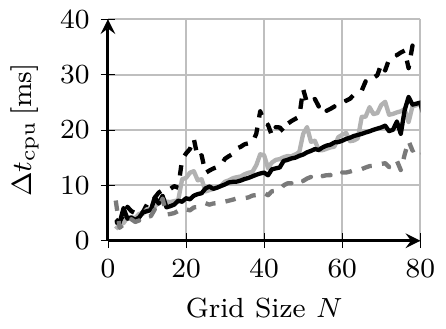}
		\caption{Van der Pol oscillator} 
		\label{fig:benchmark:open_loop:uniform_global:hermite_simpson_inc_n:vdp}
	\end{subfigure}
	\begin{subfigure}[b]{0.49\columnwidth}
		\centering
		\includegraphics{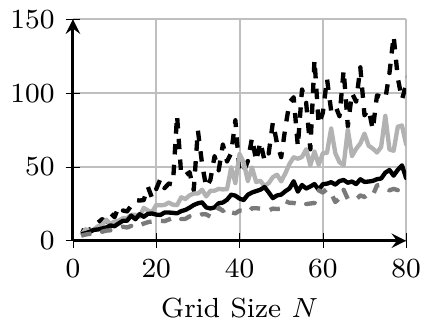}
		\caption{Rocket system} 
		\label{fig:benchmark:open_loop:uniform_global:hermite_simpson_inc_n:rocket}
	\end{subfigure}
	\caption{Computation times for selected Hermite-Simpson variants.}
	\label{fig:benchmark:open_loop:uniform_global:hermite_simpson_inc_n}
\end{figure}

A final benchmark compares the compressed and uncompressed forms as discussed in \refsec{sec:collocation} w.r.t. computation time.
\reffigc{fig:benchmark:open_loop:uniform_global:hermite_simpson_inc_n} shows the median computation times $\tcpumed$ evaluated on a PC with Ubuntu 16.04 (Intel Core i7-4770 CPU at \SI{3.4}{GHz}, \SI{8}{GB} RAM) and 20 repetitions for quadratic and constant control splines.
The compressed form has much lower computation times for both control parameterizations. Note that for generic and very large optimal control problems, \cite{kelly2017_siam} suggests the uncompressed form for a speedup, but the problem sizes are smaller in current predictive control applications and our results favor the compressed form in these cases.
The constant control representations are faster as they have less optimization parameters. 

%% file: summary.tex
\section{Conclusion}
\label{sec:summary}

Direct collocation and especially Hermite-Simpson collocation are very well suited for optimal control with variable discretization. Due to the implicit solution of the system dynamics, even smaller grid sizes are often sufficient.
There are several possibilities for selecting the control parameterization,  i.e. piecewise constant, linear mean and linear, quadratic spline parameterizations. 
The results show that increasing the degrees of freedom, for example by considering a free midpoint (linear, quadratic spline), already leads to inherent oscillations in the time-optimal solution.
This is a particularly important result for practical applications, since this "chattering by design" can stress the actuator and thus reduce its durability.
Instead, piecewise constant or linear control representations without midpoints are to be preferred for pure time-optimal control tasks.